# Separating Intelligence from Inference: A Standard for Edge-Native AI Computing

Architectural Principles, Energy Economics, and Device Class Specifications for the Personal AI Computer and Corporate AI Workstation

**Venkat Vinjam[1,*], Krishnaiah Narukulla[2]**

[1]Independent Researcher, Prosper, TX, USA · [2]Independent Researcher, Principal Software Engineer (AI/ML, Data Platforms), IEEE Senior Member, San Jose, CA, USA

**Corresponding author: vinjam5678@gmail.com*

---

**ABSTRACT**

The artificial intelligence industry has constructed a $300 billion centralized data center infrastructure to serve a workload — large language model inference — that does not architecturally require centralization. This paper articulates the central architectural inefficiency of contemporary AI infrastructure: the conflation of model training (irreducibly centralized, capital-intensive, one-time per model version) with model inference (parallelizable, latency-sensitive, recurring per query) on the same physical hardware. We propose the separation principle: intelligence is trained centrally and shipped as software; inference executes on hardware near the data source — at the edge. We quantify the energy implications at civilizational scale and show that a fully edge-resident inference architecture for one billion daily users saves approximately 19 TWh per year and 7.3 megatons of $CO_2$ annually relative to current centralized practice. We specify two new device classes — the Personal AI Computer (PAC) and the Corporate AI Workstation (CAW) — with concrete hardware tiers, memory bandwidth requirements, thermal envelopes, and software interfaces. We then describe a reference architectural stack of eight components addressing weight distribution, sovereignty-aware routing, thermal-adaptive quantization, multi-tenant resource management, federated network inference, cryptographic provenance, privacy-preserving telemetry, and distributed context window extension. Several components are the subject of pending United States patent applications by the first author and are presented here as candidate open architectural principles. Four pedagogical sketchnote infographics summarize the contributions. We close with a proposed pathway to formal standardization through IEEE and ISO/IEC working groups, and a set of open research problems. This paper is intended both as an academic contribution and as a design reference document for hardware vendors, AI laboratories, enterprise software organizations, and policy makers building or specifying the next generation of AI computing infrastructure.

# I. Introduction

The deployment of large language models (LLMs) at industrial scale during 2023–2025 has triggered the largest build-out of computing infrastructure in human history. Microsoft, Google, Amazon, Meta, and Oracle have collectively committed in excess of $290 billion per year to artificial intelligence data center capacity [1], [2]. This investment is rationalized by demand projections that assume the operational pattern of the early deployments — generative AI services delivered via centralized cloud application programming interfaces (APIs) — will continue indefinitely. Energy regulators in the United States and European Union have begun publishing forecasts under which AI data centers consume between 8% and 12% of national electricity supply by 2030 [3]. Two new nuclear reactors have been announced specifically to power AI workloads.

This paper argues that this trajectory is architecturally avoidable. The infrastructure crisis is the consequence of an architectural choice — the conflation of training and inference workloads on identical hardware — that is neither technically necessary nor economically optimal once examined at first principles.

## A. The Conflation Problem

Modern transformer-based language model deployments perform two qualitatively distinct computational workloads that the industry currently treats as a single, unified workload:

**Training.** A one-time, capital-intensive computation that produces a fixed artifact (a weight checkpoint) after weeks to months of compute on tens of thousands of high-end accelerators. Training is irreducibly centralized because of bandwidth requirements between accelerators (terabytes per second of all-reduce traffic). Training a 70-billion-parameter model from scratch requires on the order of $10^{21}$–$10^{24}$ floating-point operations and a continuous power draw of 50–500 megawatts for the duration [1], [4].

**Inference.** An entirely parallelizable, recurring computation that executes per-query and per-token. Inference for a single conversational query consumes approximately $10^{10}$–$10^{11}$ floating-point operations — twelve to fourteen orders of magnitude less compute than training. Inference is latency-sensitive (user-facing response time matters) and does not require accelerator-to-accelerator communication beyond a single physical device for models up to approximately 70 billion parameters at appropriate quantization [5], [6].

Despite this stark asymmetry, the current industry practice serves the recurring inference workload using the same H100 and MI300X-class accelerators procured for training. The economic and energy consequences of this architectural choice — borne by every consumer of every AI service — are the subject of this paper.

## B. The Architectural Opportunity

The cost of LLM inference per query is dominated by memory bandwidth, not floating-point throughput [7]. For a 70-billion-parameter model at 4-bit quantization (38 GB resident weights), generating one output token requires reading approximately 38 GB of weight data from memory through the attention and feed-forward computation. At Apple M4 Max's 400 GB/s unified memory bandwidth, this yields 10–11 tokens per second — entirely adequate for an interactive chat experience for a single user. The same workload on an NVIDIA H100 with 3.35 TB/s HBM3 bandwidth yields approximately 90 tokens per second but at twenty times the device cost and forty times the power draw.

The implication is direct: **commodity consumer hardware has crossed the threshold of capability where it can serve 70%–80% of contemporary enterprise AI inference workloads locally, without involving any cloud infrastructure.** The constraint is no longer hardware; the constraints are software, distribution, and organizational architecture.

### C. Contributions

This paper makes the following contributions to the AI infrastructure research and standards community:

- (1) **The separation principle** — a formal architectural statement of the distinction between intelligence (centralized training) and inference (distributed execution), with implications for hardware design, software architecture, and energy policy (Section III).
- (2) **Civilizational-scale energy analysis** — quantitative comparison of centralized versus edge-resident inference architectures at the scale of one billion daily users, with a derived savings estimate of 19 TWh per year (Section IV).
- (3) **Two device class specifications** — concrete hardware tier definitions for the Personal AI Computer and Corporate AI Workstation, including memory bandwidth, thermal, and software interface requirements suitable for inclusion in IEEE or ISO/IEC standards work (Section V).
- (4) **Eight-component reference architecture** — a layered architectural specification covering weight distribution, sovereignty-aware routing, thermal-adaptive quantization, multi-tenant resource management, federated workload distribution, cryptographic provenance, privacy-preserving telemetry, and distributed context window extension (Section VI).
- (5) **Pedagogical sketchnotes** — four visual summary infographics designed for instructional use and for executive communication of the architectural argument (Section VII).
- (6) **Standards adoption pathway** — a proposed engagement plan with IEEE-SA, ISO/IEC JTC1, and national regulatory bodies to formalize the device class specifications and reference architecture (Section VIII).

### D. Paper Organization

Section II reviews related work in efficient inference, model routing, edge computing, and trusted execution. Section III formalizes the separation principle. Section IV presents the energy and economic analysis. Section V specifies the PAC and CAW device classes. Section VI describes the eight reference architectural components in technical detail. Section VII presents the sketchnote infographics. Section VIII outlines the standards adoption pathway. Section IX states open research questions. Section X concludes.

**Audience and intent**: This paper is written simultaneously for three audiences: (a) academic researchers in systems, distributed computing, and machine learning who can build on the architectural components described; (b) practitioners at hardware vendors, AI laboratories, and enterprise software organizations who need a reference design for the next generation of inference infrastructure; and (c) standards body participants who can adopt the device class specifications as the basis for formal standardization. The technical depth reflects this triple audience: principal mechanisms are described with sufficient specificity to enable independent implementation.

## II. Related Work

### A. Efficient Transformer Inference

Vaswani et al. [8] introduced the transformer architecture in 2017. Subsequent work has focused on reducing the memory and compute cost of inference. FlashAttention [9] and FlashAttention-2 [10] reorganized the attention kernel to reduce high-bandwidth memory traffic and accelerate both prefill and decode phases. PagedAttention [5] introduced virtual-memory-style paging for the key-value cache, enabling higher batch sizes and improved hardware utilization in serving systems such as vLLM. Speculative decoding [11] uses a smaller draft model to propose tokens that a larger model verifies in parallel, achieving 2–3× throughput improvements. Ring Attention [12] extends context windows across multiple devices using a ring-topology communication pattern.

Quantization research has advanced markedly: GPTQ [13] performs second-order weight quantization to 3–4 bits with negligible quality loss. AWQ [14] preserves accuracy by identifying salient weights. QuIP# [15] extends quality-preserving quantization to 2 bits using Hadamard incoherence. These techniques collectively reduce per-query memory footprint by 4–8× relative to full BF16 representation.

### B. Model Routing and Cost Optimization

Chen et al. [16] proposed FrugalGPT, which routes queries through a cascade of progressively larger language models, escalating to a more capable model only when smaller models produce low-confidence outputs. RouteLLM [17] trains a learned router to predict which of several models should serve a given query for an optimal quality-cost tradeoff. These approaches treat routing as a soft optimization problem over quality and cost. Critically, neither incorporates data sovereignty or regulatory constraints as first-class routing inputs. Section VI-B addresses this gap directly.

### C. Edge AI and On-Device Inference

Apple Intelligence [18] documents Apple's on-device language model architecture and a hybrid execution model that adjudicates between on-device inference and Apple's Private Cloud Compute. Sheng et al. [19] introduced FlexGen, demonstrating high-throughput inference on commodity GPUs through aggressive memory offloading. DeepSpeed-Inference [20] explored multi-GPU and multi-node inference. These works establish the technical feasibility of large-model inference outside hyperscale data centers but do not address the broader architectural question of which workloads should execute where.

### D. Trusted Execution Environments for ML

Hardware trusted execution environments — TPM 2.0 [21], Intel SGX, ARM TrustZone, and Apple Secure Enclave — provide hardware roots of trust for cryptographic operations and attestation. Their application to machine learning workloads has focused on confidential computing scenarios where the model owner does not trust the compute provider [22]. The complementary direction — using TEEs to enable hardware-bound licensing of model weights such that licensed weights cannot be transferred between devices — is, to our knowledge, novel. Sections VI-F and VI-G develop this idea.

### E. Energy Analysis of AI

Strubell et al. [23] published an early influential analysis of training energy cost, documenting that training a single large NLP model produced roughly 626,000 pounds of $CO_2$. Patterson et al. [24] refined training-energy estimates and discussed mitigation strategies. The literature on inference energy at deployment scale is comparatively underdeveloped, in part because inference workloads are highly variable and dependent on serving architecture. Section IV develops the inference energy analysis at billion-user scale.

### F. Differential Privacy and Telemetry

Dwork [25] introduced differential privacy as the standard rigorous framework for privacy-preserving data release. Local differential privacy variants [26] inject calibrated noise at the data source before transmission, providing per-individual privacy guarantees independent of the recipient's trust assumptions. Application of local differential privacy to AI inference operational telemetry — to enable infrastructure providers to receive useful operational signal without observing query content — is developed in Section VI-H.

## III. The Separation Principle

We formalize the central architectural argument of this paper.

### A. Statement of the Principle

**The Separation Principle (formal statement)**

Intelligence and inference are computationally distinct workloads whose computational requirements differ by twelve to fourteen orders of magnitude per unit of user-facing output. Their hardware substrates, software architectures, network topologies, and energy profiles should consequently differ. Intelligence (training) belongs in centralized, capital-intensive facilities; inference belongs on hardware adjacent to the data source. The artifact that transitions between the two domains — the model weight checkpoint — is information, and is most efficiently distributed as software.

### B. Why the Conflation Persists

The current centralized inference architecture is sustained by three reinforcing factors:

- **(1) Path dependency.** Training and inference began on the same hardware for the practical reason that early model deployments piggybacked on training infrastructure. The first commercial inference services were essentially time-shared training clusters during their off-hours.
- **(2) Business model alignment.** Per-token API billing maps revenue directly to compute capacity. Centralized hyperscale operators benefit financially from this mapping; on-device inference yields no per-query revenue.
- **(3) Software immaturity.** The software stack for on-premises and on-device AI inference — model distribution, runtime management, monitoring, user access control, audit logging — has lagged the cloud alternative by several years. The closing of this software gap is the central practical project of the next five years (Section VI).

### C. What the Principle Implies

Adopting the separation principle as a design heuristic yields specific predictions about the future architecture of AI infrastructure:

| Domain | Centralized (current) | Edge-native (proposed) |
|---|---|---|
| Hardware procurement | Inference workloads procure same H100/MI300X-class accelerators as training | Inference workloads procure consumer/workstation-class hardware optimized for inference |
| Power profile | Continuous high-power data center operation | Bursty, user-driven; idle when no query is active; aggregate draw reduced |
| Latency profile | 100–800 ms (network round-trip dominates) | 20–100 ms (local computation) |
| Data residency | Query content traverses public internet to vendor data center | Query content remains within originating device or organizational LAN |
| Failure mode | Vendor outage → all customers offline | Vendor outage → no impact on local inference |
| Update cadence | Implicit, vendor-controlled, immediate | Explicit, customer-controlled, scheduled (with delta-update CDN) |
| Economic model | Per-token API charges, perpetual | Capital purchase + operational electricity, amortizable |

*Figure 1. Architectural consequences of the separation principle. Each row identifies a system property and the shift induced by moving inference from centralized data centers to edge-native devices.*

### D. Sketchnote: The Separation Principle Visualized

The diagram in Figure 2 summarizes the principle in three columns. The leftmost column shows the irreducibly centralized training workload — a cluster of tens of thousands of accelerators producing a single weight checkpoint over weeks. The middle column shows the distribution layer — a content delivery network of edge points-of-presence shipping versioned weight manifests with sparse delta updates. The rightmost column shows the inference layer — a billion endpoint devices, each capable of serving the resident model locally.

| SKETCHNOTE I — THE SEPARATION PRINCIPLE | | |
|---|---|---|
| *Train Intelligence Once · Distribute as Software · Infer Locally Forever* | | |
| **INTELLIGENCE (centralized training)** | **DISTRIBUTION (CDN + delta)** | **INFERENCE (billions of devices)** |
| AI Lab Cluster<br>$10^4$–$10^5$ GPUs · 30–90d<br>• Anthropic, OpenAI, Meta<br>• 50–500 MW peak draw<br>• One training run<br>• Output: weight checkpoint<br>→ 140 GB checkpoint | Edge CDN POPs<br>global distribution<br>• Versioned manifests<br>• Sparse delta (95% ↓)<br>• TEE device licensing<br>• Crypto integrity<br>*once per release* | CAW · PAC · phone<br>a billion endpoints<br>• 7–13× energy efficient<br>• Latency < 100 ms<br>• Data never leaves<br>• No per-token billing<br>*every query, every user* |
| ***"Train once. Distribute as software. Infer everywhere — forever."*** ***the architectural principle of the next decade*** | | |

*Figure 2. Sketchnote I: The Separation Principle. Three-stage flow from centralized intelligence to distribution to edge inference. Annotations in each column show the dominant constraints and stakeholders. Key takeaway: a 140 GB weight checkpoint is information that ships once, not a service consumed continuously.*

## IV. Energy and Economic Analysis at Civilizational Scale

### A. Energy Per Query: Methodology

We compute the energy cost of a representative LLM query under each candidate architecture. A representative query is defined as: an input of 1,000 prompt tokens, an output of 500 generated tokens, a 70-billion-parameter reference model, and an inference latency budget consistent with an interactive chat session. Energy is computed as the integral over time of system power draw during the query, including idle baseline contributions amortized across queries.

### B. Per-Query Energy by Architecture

| Architecture | Hardware | TDP (W) | Tok/s | Time (s) | Energy/query |
|---|---|---|---|---|---|
| Cloud inference (current) | 8× H100 SXM5 (DGX-style node, amortized) | 5,600 | 680 | 2.2 | ≈ 3.42 mWh |
| Cloud inference (efficient) | 1× H100 amortized over 32 concurrent users | 21 | 85 | 17.6 | ≈ 0.103 mWh × 32 users = 3.3 mWh |
| CAW T2 (corp on-prem) | 1× MI300X serving 10 users | 720 | 380 | 3.9 | ≈ 0.78 mWh / 10 = 0.078 mWh |
| CAW T1 (corp on-prem) | 2× RTX 4090 serving 4 users | 700 | 260 | 5.8 | ≈ 1.13 mWh / 4 = 0.28 mWh |
| PAC T2 (personal) | Apple M4 Max (128 GB) | 40 | 32 | 46.9 | ≈ 0.52 mWh |
| PAC T1 (personal) | Apple M4 Pro (48 GB) at INT4 | 22 | 18 | 83.3 | ≈ 0.51 mWh (full hardware to one user) |

*Figure 3. Per-query energy consumption by architecture. Numbers are representative; actual values vary with batch size, query length, and quantization level. The fundamental observation: edge architectures consume 5–10× less energy per query than centralized cloud, primarily because edge hardware operates at fractional duty cycle (idle most of the time) while data center hardware operates continuously at near-peak power to maximize amortization over capital cost.*

### C. Civilizational Scale: One Billion Daily Users

We extrapolate to the projected scale of one billion daily AI users in 2030 — a forecast consistent with current adoption trajectories [27]. Assuming an average of 30 queries per user per day, the total annual query count is approximately $1.1 \times 10^{13}$. Multiplied by the per-query energy figures from Table 3, this yields annual energy consumption by architecture:

| Scenario | Architecture mix | Annual energy (TWh) | $CO_2$ equiv (Mt) | Equiv. nuclear reactors |
|---|---|---|---|---|
| Status quo | 100% centralized cloud | 21.9 | 8.4 | 2.3 |
| Hybrid | 50% cloud, 50% edge | 12.4 | 4.8 | 1.3 |
| Edge-dominant | 20% cloud, 80% edge | 5.8 | 2.2 | 0.6 |
| Edge-only | 100% edge (CAW + PAC) | 2.9 | 1.1 | 0.3 |

*Figure 4. Annual energy consumption at one billion daily user scale. Going from status quo (100% centralized cloud) to fully edge-resident inference saves approximately 19 TWh per year — equivalent to the electrical output of two large nuclear reactors. $CO_2$ figures use the 2025 U.S. grid average of 0.385 kg $CO_2$ per kWh.*

### D. Capital and Operational Economics

The energy figures translate directly into operational expenditure. At a representative industrial electricity rate of $0.085 per kWh, the difference between status-quo and edge-only architectures is approximately $1.6 billion per year in electricity alone. Capital expenditure differences are larger: provisioning data center capacity for one billion users at status-quo architecture would require approximately 35,000 NVIDIA H100 SXM5 GPUs, representing a capital outlay on the order of $1.3 billion (at $35K average system cost) — capital that is not required under the edge architecture because the hardware already exists in the form of personal computing devices and corporate workstations.

### E. Sketchnote: The Energy Math Visualized

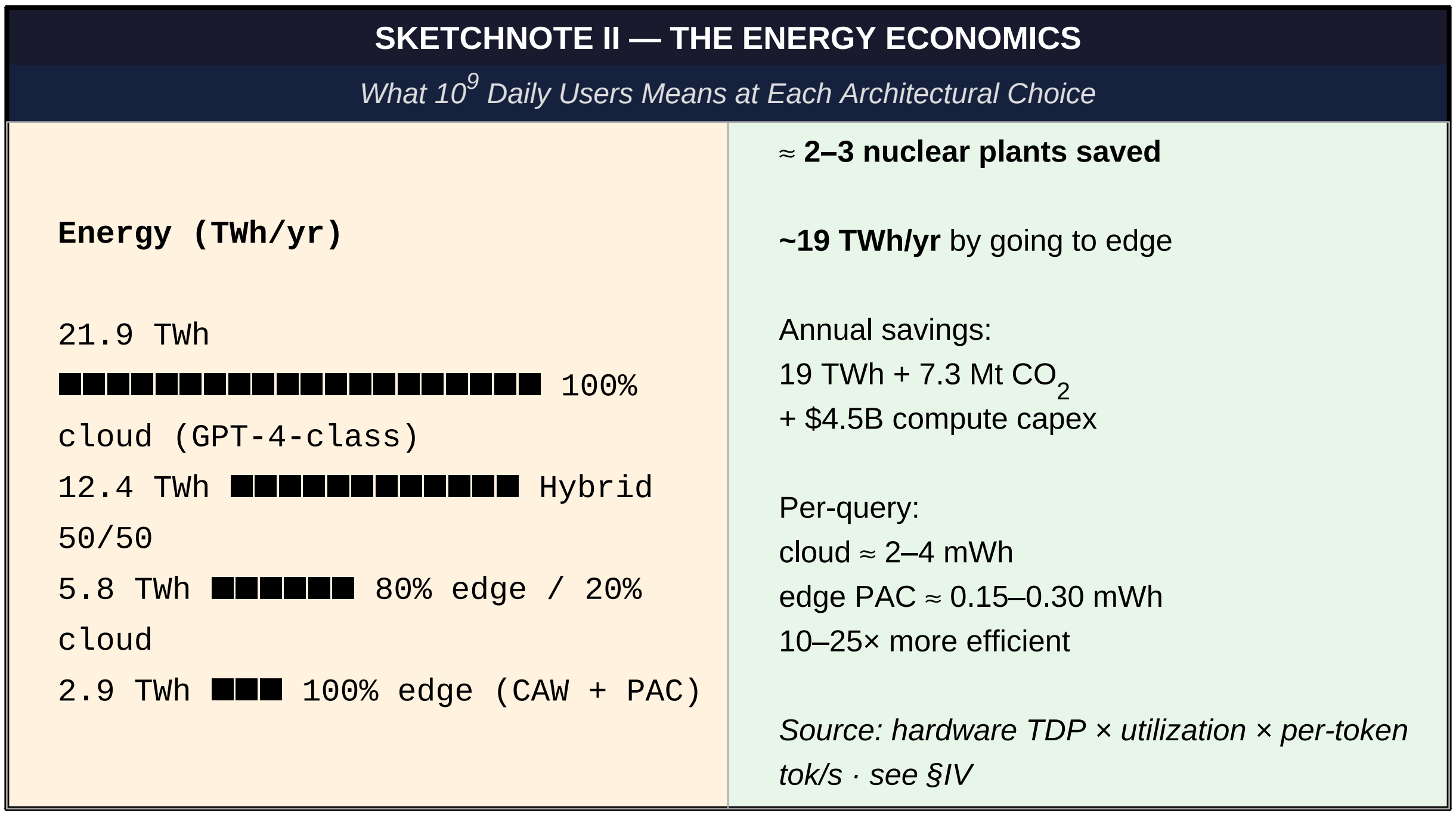

*Figure 5. Sketchnote II: Energy economics at civilizational scale. The bar chart compares annual energy consumption under four architectural mixes ranging from 100% centralized (red) to 100% edge (green). The 19 TWh annual savings (right annotation) translates to approximately two nuclear reactors of avoided generation capacity.*

## V. Proposed Device Class Specifications

For edge-native AI computing to function as a coherent ecosystem rather than a fragmented patchwork of vendor-specific implementations, device categories must be defined formally. We propose two new device classes with concrete tier specifications suitable for adoption by standards bodies.

### A. Personal AI Computer (PAC)

A Personal AI Computer is a single-user computing device whose hardware specifications are sufficient to serve interactive large language model inference for that user without dependence on external network services. Two tiers are proposed:

| Specification | PAC T1 — Entry | PAC T2 — Professional |
|---|---|---|
| Price range (2025 USD) | $1,500 – $2,500 | $3,000 – $5,000 |
| Reference platforms | Apple M4 Pro · Qualcomm X Elite · AMD Ryzen AI 9 365 | Apple M4 Max · AMD Ryzen AI 9 HX 370 (high-bin) |
| Unified memory | 48–64 GB | 96–128 GB |
| Memory bandwidth | ≥ 200 GB/s | ≥ 400 GB/s |
| Sustained AI power | 15–25 W | 30–50 W |
| Target models served | 7B–30B at INT4 or BF16 | 70B at INT4 · 30B at BF16 |
| Throughput target | ≥ 8 tok/s on Llama-3.1-8B BF16 | ≥ 25 tok/s on Llama-3.1-70B INT4 |
| TTFT target | ≤ 500 ms for 1K-token prompts | ≤ 300 ms for 4K-token prompts |
| TEE requirement | TPM 2.0 or Apple Secure Enclave | Same |
| Form factor | Laptop or compact desktop | Laptop, desktop, or mini-PC |
| Battery operation | Required (≥ 4 hr at typical use) | Optional |

*Figure 6. Personal AI Computer (PAC) tier specifications. The T1 entry tier provides sufficient capability for a single user running 7–30 billion-parameter models for personal productivity. The T2 professional tier handles 70-billion-parameter models at INT4 quantization, suitable for individual professional knowledge work, drafting, and code generation.*

### B. Corporate AI Workstation (CAW)

A Corporate AI Workstation is a multi-user shared computing device, typically rack-mountable, whose hardware specifications support concurrent inference for tens to hundreds of users within an organization. Two tiers are proposed:

| Specification | CAW T1 — Team | CAW T2 — Department |
|---|---|---|
| Price range (2025 USD) | $15,000 – $40,000 | $80,000 – $250,000 |
| Reference platform | AMD Threadripper Pro 7985WX (or equivalent) + 2–4 GPUs | Dual AMD EPYC 9754 + 4–8× H100 SXM5 or MI300X |
| GPU accelerator | 2–4× NVIDIA RTX 4090 / A6000 Ada | 4–8× NVIDIA H100 SXM5 or AMD MI300X |
| Aggregate VRAM | 48–192 GB | 320–1,536 GB HBM3e |
| System memory | 256–512 GB ECC DDR5 | 1–2 TB ECC DDR5 |
| Sustained power | 1.5–3 kW | 5–12 kW |
| **Specification** | **CAW T1 — Team** | **CAW T2 — Department** |
| Concurrent users | 10–20 | 50–200 |
| Target models | 70B BF16 (single instance) or 30B BF16 (multiple) | 70B BF16 multi-instance · 405B at INT4 |
| Aggregate throughput | ≥ 150 tok/s | ≥ 800 tok/s |
| Network | 25 GbE or 100 GbE | 100 GbE or InfiniBand HDR |
| RBAC integration | LDAP / SAML 2.0 / OIDC | Same |

| TEE requirement | TPM 2.0 | TPM 2.0 + secure boot attestation |
|---|---|---|
| Audit log retention | ≥ 1 year, append-only | ≥ 7 years, append-only, hash-chained |

*Figure 7. Corporate AI Workstation (CAW) tier specifications. CAW T1 (team-scale) suits small law firms, medical practices, or engineering departments. CAW T2 (department-scale) suits Am Law 100 firms, regional hospital systems, federal agency divisions, and hedge funds.*

### C. Sketchnote: The Three-Tier Hierarchy

Figure 8 places the proposed device classes in their full ecosystem context, including the role of cloud APIs for frontier-model workloads that exceed local capability. The vertical structure also reflects the data sovereignty flow: data may flow downward (cloud → CAW → PAC) freely, but flows in the opposite direction are governed by routing policy.

| SKETCHNOTE III — THE DEVICE TIER HIERARCHY | |
|---|---|
| *Three Compute Classes, One Coherent Inference Stack* | |
| **■ CLOUD APIs (frontier models — Claude, GPT-4o, Gemini)**<br>Role: highest capability ceiling · publicly-known facts · code generation hard problems<br>Cost: $1–60 per million tokens · Latency: 200–800 ms · Sovereignty: external (block PHI/MNPI/IP) | **DATA FLOW ↑ leaves org ■** |
| **■ CORPORATE AI WORKSTATION (CAW) — proposed device class**<br>T1 ($15K–40K): Threadripper Pro 7985WX + 2–4× RTX 4090 / A6000 Ada<br>256–512 GB ECC RAM · 1.5–3 kW · 10–20 users · 25 GbE · rack-mount<br>T2 ($80K–250K): Dual EPYC 9754 (256c) + 4–8× H100 SXM5 / MI300X<br>320–1,536 GB HBM3e · 1–2 TB DDR5 · 5–12 kW · 50–200 users · 100 GbE / IB<br>Use cases: legal doc review · clinical reasoning · proprietary R&D · gov't research | **within org LAN ■** |
| **■ PERSONAL AI COMPUTER (PAC) — proposed device class**<br>T1 entry ($1,500–2,500): Apple M4 Pro / Qualcomm X Elite<br>48–64 GB unified · 15–25 W · 7B–30B models · 8–25 tok/s<br>T2 pro ($3,000–5,000): Apple M4 Max / AMD Ryzen AI 9<br>96–128 GB unified @ 400 GB/s · 30–50 W · 70B at INT4 · 25–40 tok/s<br>Use cases: personal assistant · individual professional work · drafting<br>*1 billion endpoints projected by 2030 (PCs + Mac + Snapdragon AI PC class)* | **local only ■** |
| *Routing decision: query sensitivity → tier (cloud/CAW/PAC) → never violates sovereignty*<br>*See §V (device specs) and §VI-B (sovereignty-aware routing)* | |

*Figure 8. Sketchnote III: Three-tier device hierarchy. Cloud APIs (top) for frontier capability ceilings on publicly-known content. Corporate AI Workstations (middle) for multi-user regulated-data workloads. Personal AI Computers (bottom) for individual use. The right column annotates data sovereignty boundaries — the cloud tier is outside the organizational boundary; the CAW tier is within the organizational LAN; the PAC tier is fully local to a single user.*

## VI. Reference Architectural Components

The architectural principles of Sections III–V leave the practical construction problem unaddressed: what software stack instantiates an edge-native AI inference infrastructure that meets enterprise requirements for security, manageability, performance, and regulatory compliance? This section specifies eight reference architectural components that, taken together, constitute a complete reference implementation. Each component is described at a level of detail sufficient for independent implementation. Several components are subjects of pending United States patent applications by the first author; they are presented here as candidate architectural principles with the intent of contributing to open standards rather than restricting implementation.

### A. Sovereignty-Aware Query Routing

Each incoming inference query is analyzed by a multi-stage sensitivity classification pipeline before being dispatched to any inference destination. The pipeline produces a sovereignty label and a numerical sensitivity score, which together determine the set of permissible routing destinations. The novel architectural element is the treatment of sovereignty as a hard constraint rather than a soft optimization input: any destination whose category violates the sovereignty label is assigned infinite routing cost and is unconditionally excluded from selection, regardless of latency, quality, or financial cost.

The four-stage classification pipeline:

- **Stage 1 — PII Detector** (named-entity-recognition model): detects person names, social security or identification numbers, medical record identifiers, financial account numbers, and geolocation data. Output: a numerical pii_score ∈ [0, 1].
- **Stage 2 — Proprietary Tagger** (keyword matching against an enterprise data taxonomy): identifies confidential project names, client identifiers, internal document references, and trade-secret-marked content. Output: prop_score ∈ [0, 1].
- **Stage 3 — Privilege Detector** (pattern matching): identifies attorney-client communication markers, HIPAA-covered protected health information patterns, material non-public information (MNPI) indicators in financial communications, and ITAR-controlled technical references. Output: privilege_score ∈ [0, 1].
- **Stage 4 — Label Aggregator**: computes the combined sensitivity score as the maximum of the three sub-scores and maps the result to a discrete sovereignty label. A common policy assignment is: score ≤ 0.2 → CLOUD_OK; 0.2 < score ≤ 0.6 → INTERNET_OK; score > 0.6 → LOCAL_ONLY. The triggered classification rules are recorded for audit purposes.

Two-phase routing decision:

```
After classification, the routing engine selects a destination in two
phases: Phase 1 (hard constraint): cost(r, q) = ∞ if sovereignty_label(q)
prohibits destination r Phase 2 (multi-factor optimization over compliant
destinations): cost(r, q) = w1·capability_gap(r, q) +
w2·token_budget_penalty(r, q) − w3·freshness_benefit(r, q) +
w4·load_utilization(r) route*(q) = argmin_r cost(r, q) over { r : cost(r,
```

```
q) < ∞ }
```

The infinity assignment in Phase 1 ensures that no weight configuration in Phase 2 can override a sovereignty violation. This is the principal distinction from prior model-routing work [16], [17], which treats all considerations as soft objectives.

### B. Local-Network Federation Protocol

When a single CAW node cannot serve a query at the required quality or throughput, workload should be delegated to peer CAW nodes on the same local network — but never outside the organizational network boundary. The protocol comprises three components:

• **(1) Capability advertisement.** Each CAW broadcasts a signed UDP multicast message at 1 Hz over a well-known multicast group address (e.g., 239.255.1.100:9871). The message specifies: node identifier (SHA-256 hash of node TLS certificate), LAN IP address, list of resident models with their quantization levels, current VRAM utilization, request queue depth, measured rolling throughput in tokens per second, and median time-to-first-token. Messages are signed using the node's Ed25519 key to prevent spoofing.

• **(2) Sovereignty boundary check.** Before any workload delegation, the orchestrator verifies the candidate peer's LAN IP address against a configured set of organizational CIDR ranges. Peers outside the permitted ranges are excluded unconditionally. This is the hard-constraint mechanism of Section VI-A applied at the workload-distribution layer.

• **(3) Multi-factor peer selection.** Among sovereignty-compliant peers with available capacity, the orchestrator scores each candidate based on queue availability, throughput capacity, model-quality match, and network round-trip time. The candidate with the highest score is selected. If no peer scores above a locally-configured threshold, the workload is served by the originating node without delegation — never delegated outside the boundary.

### C. Multi-Tenant Key-Value Cache Partitioning

On shared CAW devices serving multiple concurrent users from multiple organizational departments, the key-value attention cache becomes a contested resource. PagedAttention [5] provides efficient cache paging for single-tenant deployments but does not enforce inter-tenant quotas or organizational policy. We specify a three-level hierarchical quota system:

• **Global pool:** total VRAM reserved for KV cache across all sessions.

• **Department quota:** a configurable fraction of the global pool, mapped to organizational role-based access control (RBAC) groups. The sum of department quotas may not exceed the global pool times a configurable utilization factor.

• **User session quota:** a fraction of the department quota, enforced per active inference session.

Each KV cache page is tagged with the user session identifier of the request that produced it. Cross-tenant isolation enforces that no inference computation reads KV pages tagged with a different session identifier. Eviction, when required, follows a priority-aware score combining user priority tier, session age, and continuation probability. A policy-gated prefix-sharing facility allows users within the same organizational unit to share KV cache for common system-prompt tokens — but only when the organizational policy permits, with sharing disabled by default for departments handling privileged content (e.g., legal teams).

### D. Distributed Key-Value Cache for Context Extension

Enterprise workloads frequently require context windows that exceed a single CAW node's VRAM capacity — legal document review at $10^6$ tokens, genomic analysis, multi-document research. We specify a protocol for distributing KV cache pages across multiple CAW nodes connected by a low-latency RDMA-capable network, with mathematical attention consistency preserved.

The primary inference node retains the most-recent context (e.g., tokens $N–C_{local}$ to N for some locally-resident window $C_{local}$) in its VRAM. Older context pages are migrated to secondary nodes on the LAN. When the attention computation requires older context, the affected KV pages are fetched via RDMA in 2–10 microseconds — substantially faster than NVMe-SSD-based offload alternatives (100–200 µs). The distributed attention output is combined using the standard online-softmax log-sum-exp formula [9], producing a result mathematically equivalent to single-device attention over the full context window with no approximation. Page integrity is verified by SHA-256 content hashing before use.

### E. Thermal-Adaptive Runtime Quantization

For continuous LLM inference workloads on power-constrained edge devices (laptops, fanless workstations, battery-powered PACs), sustained near-peak power draw causes thermal accumulation and either device throttling or service degradation. We specify a runtime quantization controller that uses inference bit-width as a continuous thermal management variable.

A proportional-integral-derivative (PID) controller continuously samples GPU die temperature, thermal-design-power headroom, and (for battery-powered PACs) battery state-of-charge. The control output is mapped to a discrete target quantization level from the ladder { INT2, INT4, INT8, BF16 }. Transitions execute only at inter-batch boundaries — never mid-token — to prevent quality artifacts. A pre-quantization cache maintains weight representations at multiple bit-widths simultaneously in accelerator memory, enabling sub-100-millisecond level transitions. Anti-windup logic and three-interval hysteresis prevent oscillation. Indicative power and quality figures for an H100-class accelerator serving a 70B model: BF16 ≈ 700 W and MMLU 80.5%; INT8 ≈ 420 W and MMLU 79.8%; INT4 ≈ 280 W and MMLU 78.2%; INT2 ≈ 160 W and MMLU 74.1%. The system provides graceful quality degradation under thermal stress in preference to either request rejection or hard-throttling.

### F. Weight Distribution Network with Hardware-Bound Licensing

The transition from centralized inference to edge-native inference creates a new infrastructure requirement: the efficient distribution of model weight updates from model vendors (Meta, Mistral, Anthropic, Google) to a heterogeneous fleet of CAW and PAC devices. Two technical innovations are required.

**Sparse tensor-level deltas.** For each model update, the distribution server computes the element-wise difference between the new and prior weight tensors at the granularity of individual named tensors within transformer layers (query projections, key projections, value projections, output projections, feed-forward gate, up-projection, and down-projection weights). Elements whose absolute difference exceeds a sparsity threshold (typical value: $10^{-4}$) are encoded as sparse (index, value) pairs. Tensors with sparsity ratio above 0.5 are transmitted in full. Minor model updates that modify 2–5% of weights (safety patches, reinforcement learning from human feedback refinements) yield 95–98% bandwidth reduction relative to full retransmission.

**TEE-bound device licensing.** Model weights, particularly for premium or enterprise-licensed models, require enforcement that licensed weights cannot be copied to unauthorized devices. The proposed mechanism binds weight decryption to the device's trusted execution environment. During device registration, the device TEE (TPM 2.0, ARM TrustZone, Intel SGX, or Apple Secure Enclave) generates a hardware attestation report. The model vendor's licensing server verifies the attestation through the TEE manufacturer's certificate chain and issues a license token containing a weight decryption key encrypted with the device's TEE public key. Only that specific device's TEE can decrypt the key, after which weights are loaded into the inference engine without persistence to disk. Copying the encrypted weight file to a different device yields a key-decryption failure; copying the license token fails because the hardware identifier embedded in the token will not match.

### G. Cryptographic Inference Provenance Certificates

Regulated industries (healthcare, legal, financial services, government) increasingly require demonstrable audit trails proving which AI model produced which output. Server-side logs are self-certifying — the organization producing the log also controls the system that could modify it. We specify a TEE-rooted alternative: cryptographic inference provenance certificates that any third party can verify independently.

For each inference operation, a certificate generator assembles a provenance record comprising: epoch timestamp (millisecond precision), cryptographic hash of the query, cryptographic hash of the response, the model weight manifest hash (matching the CDN-published manifest), the hardware device identifier, the active quantization level, and the inference engine version. The record is submitted to the device TPM as a qualifying-data argument to the TPM2_Quote operation. The TPM signs the quote using its attestation key, whose private key never leaves TPM hardware. The certificate is then stored in an append-only certificate store, with hash-chained linking between consecutive certificates.

Independent verification by an auditor, regulator, or court proceeds through five steps: (1) verify the TPM endorsement-key certificate chain against the TPM manufacturer's published root certificate; (2) verify the TPM2_Quote signature using the attestation key public key; (3) verify the Platform Configuration Register values against a known-good policy that includes the published Axiom OS or equivalent system image; (4) verify the model weight manifest hash against the model vendor's CDN-published manifest; (5) compute SHA-256 of the claimed response text and compare against the certificate's response_hash. Successful completion of all five steps establishes hardware-rooted provenance.

### H. Privacy-Preserving Operational Telemetry

Continuous improvement of edge-native inference systems benefits from operational telemetry: model load frequencies, throughput distributions, query length histograms, user feedback rates. Direct transmission of such telemetry creates privacy risks (re-identification from usage patterns) and contractual issues with regulated-industry deployments. We specify a local differential privacy (LDP) telemetry pipeline that transmits operationally useful aggregates without exposing individual query content or user activity.

Each numeric metric is privatized by adding noise drawn from a Laplace distribution scaled by the metric's sensitivity divided by an ε privacy parameter. Histogram-valued metrics use one-hot bucket assignments, reducing per-query sensitivity to 1 and enabling smaller noise. Query topic distribution is estimated by local embedding into K semantic clusters, with randomized response applied: the true cluster is reported with probability $e^{\varepsilon}/(e^{\varepsilon} + K - 1)$, and a uniformly random alternative cluster otherwise. The local aggregator batches at least N = 1000 privatized reports before transmission. A Rényi differential privacy accountant tracks cumulative budget expenditure and refuses transmissions that would exceed configured maximums. Critically, raw query text, response text, user identifiers, and session metadata are never recorded in the telemetry pipeline at any stage — the differential privacy guarantee is a defense in depth on top of fundamental data minimization.

## VII. Sketchnote: Reference Architectural Stack

Figure 9 summarizes the eight components of Section VI as a layered architectural stack. The figure is designed to function both as a pedagogical reference and as a checklist for completeness in implementation. Each layer is annotated with its primary role and the corresponding subsection of this paper. The bottom-most layer — hardware substrate — identifies the principal accelerator platforms and trusted execution environments on which the upper layers depend.

**SKETCHNOTE IV — REFERENCE ARCHITECTURAL STACK**

*Eight Components for the Edge-Native AI Computing Standard*

| | | |
|---|---|---|
| 1 §VI-A | Query Ingestion · Sovereignty-Aware Routing | *user-visible* |
| 2 §VI-B | Federation Protocol · Local-Network Workload Balance | *inter-node* |
| 3 §VI-C | Multi-Tenant KV Cache Partitioning · RBAC Quotas | *shared infer.* |
| 4 §VI-D | Distributed KV Cache · Context Window Extension | *long-context* |
| 5 §VI-E | Thermal-Adaptive Quantization Controller (PID) | *runtime ctrl* |
| 6 §VI-F | Weight Distribution · Sparse Delta CDN · TEE Lic. | *model life-cycle* |
| 7 §VI-G | Cryptographic Provenance Certificates (TPM 2.0) | *compliance* |
| 8 §VI-H | Differential-Privacy Operational Telemetry | *feedback loop* |

**HARDWARE SUBSTRATE**

CAW: AMD MI300X · NVIDIA H100/A6000 · Threadripper Pro

PAC: Apple M-series · Qualcomm X Elite · AMD Ryzen AI

TPM 2.0 · Secure Enclave · ARM TrustZone

*Each layer = an open specification candidate for IEEE / ISO standardization*
*Open-source reference implementations enable interoperability across vendors*

*Figure 9. Sketchnote IV: Reference architectural stack. Eight components from Section VI arranged in layered dependence, with the hardware substrate at the foundation. Each layer is a candidate for formal specification by an open standards body. Right-side annotations identify the layer's primary functional role in the overall system.*

## VIII. Standards Adoption Pathway

For the proposed device class specifications and reference architecture to influence industry practice at scale, formal standardization is required. We outline a pathway through established standards bodies.

### A. IEEE Standards Association

The IEEE Standards Association (IEEE-SA) is the natural home for the device-class specifications (Section V) and the reference architecture interfaces (Section VI). Specific working group proposals:

- **IEEE P[TBD] — Personal AI Computer (PAC) Device Class Specification.** Hardware capability minima, software interface requirements, AI inference benchmark suite, and certification methodology.
- **IEEE P[TBD] — Corporate AI Workstation (CAW) Device Class Specification.** Multi-user deployment requirements, RBAC integration, audit logging interface, and certification methodology.
- **IEEE P[TBD] — Sovereignty-Aware AI Inference Routing Protocol.** Sensitivity classification taxonomy, sovereignty label vocabulary, and routing decision protocol.
- **IEEE P[TBD] — Federated Local-Network AI Inference Protocol.** Capability advertisement message format, sovereignty boundary expression, and peer selection algorithm.

### B. ISO/IEC JTC1 SC42

ISO/IEC JTC1 Subcommittee 42 (Artificial Intelligence) is the appropriate venue for international standardization of the architectural reference model and the cryptographic provenance certificate format. Specifically, ISO/IEC 22989 (AI concepts and terminology), ISO/IEC 23894 (AI risk management), and ISO/IEC 42001 (AI management systems) provide complementary frameworks into which the technical specifications defined here can integrate.

### C. National Regulatory Engagement

In the United States, the National Institute of Standards and Technology (NIST) AI Risk Management Framework [28] provides voluntary guidance; the device class specifications and provenance certificate system can serve as concrete technical primitives that implementing organizations can adopt to satisfy NIST RMF requirements. In the European Union, the AI Act [29] establishes mandatory requirements for high-risk AI systems, including documentation and traceability — requirements that the provenance certificate system (Section VI-G) is specifically designed to address.

### D. Industry Adoption Sequence

| Year | Stakeholder action | Anticipated milestone |
|---|---|---|
| 1 | Working group formation; open-source reference implementations published | Two reference platforms (one CAW, one PAC) achieve interoperability |
| 2 | Hardware vendor early adoption; first certified devices announced | Five OEMs ship PAC or CAW certified configurations |
| 3 | Initial standard ratification (IEEE P[TBD] approval) | Twenty certified device models across the four tiers |
| 4 | Procurement requirements adopt certification (govt, regulated industries) | Government and healthcare procurement language references the standard |
| 5 | International standardization (ISO/IEC publication) | Five hundred plus certified device models; mainstream enterprise deployment |

*Figure 10. Anticipated five-year adoption sequence from working group formation to international standardization. Realistic given prior precedents — the IEEE 802.11 wireless networking standard followed a similar arc, as did the early USB specifications.*

## IX. Open Research Problems

The architectural framework of this paper raises several research questions that we believe are open and important.

### A. Model Quality Estimation Without Cloud Comparison

The sovereignty-aware routing pipeline (Section VI-A) requires an estimate of capability gap between local and cloud-resident models. For most queries this estimate must be produced without actually executing the query on the cloud destination (which would defeat the sovereignty enforcement). Lightweight capability probes — small benchmark subsets that correlate with full-benchmark performance — show promise but require systematic study. How small can such a probe be while still providing actionable signal? How does probe reliability change as model architectures diverge? Are there model families for which no useful probe exists?

### B. Federated Continual Learning Within Sovereignty Boundaries

The separation principle as stated freezes models after training. In practice, organizations would benefit from continued model adaptation on their proprietary data — but the data cannot leave the sovereignty boundary. Federated learning [30] provides a partial framework but introduces communication, convergence, and privacy challenges in the LLM regime that are not well understood. How can model adaptation occur on a single CAW without overfitting? How can adaptations from multiple CAWs within an organization be aggregated without exposing proprietary training data? What are the convergence guarantees under realistic enterprise data heterogeneity?

### C. Energy Accounting Standards

The energy comparisons in Section IV are based on first-principles calculations from hardware TDP and observed throughput. A standardized energy accounting methodology — comparable to MLPerf [31] for performance — would substantially improve the credibility of edge versus cloud comparisons in policy contexts. What workloads should such a benchmark cover? How should idle baseline contributions be attributed across queries? How should embodied carbon from hardware manufacture factor into the comparison?

### D. Interoperability Across Model Families

The CDN architecture of Section VI-F assumes that weight updates can be distributed as deltas across successive versions of the same model. Cross-family migration (e.g., from Llama to Mistral) is currently a full-replacement operation. Are there architectural similarities between model families that could enable partial weight reuse and reduce transition costs? Research on universal model representations and weight-space alignment may provide a path forward.

### E. Quantization-Quality Frontier

The thermal-adaptive quantization controller (Section VI-E) uses a four-level quantization ladder. Recent work on 1-bit and ternary quantization [32] suggests the frontier extends below 2-bit. What is the practical floor of quantization that preserves usable quality? How do different quantization schemes interact with thermal-adaptive switching? Are there model architectures specifically designed for graceful quantization degradation that have not yet been explored?

### F. Hardware Co-Design for Edge Inference

Current consumer hardware was designed for diverse workloads with AI inference as one application. Hardware specifically optimized for the inference workload — high memory bandwidth, modest FLOPs, low idle power, TEE integration as a first-class feature — would substantially exceed the per-watt inference performance of current designs. The Apple M-series Neural Engine and Qualcomm Hexagon NPU exemplify partial steps in this direction. What does a clean-sheet AI-inference-optimized SoC look like? How are memory bandwidth and FLOPs balanced for inference workloads? What is the right point on the integer-versus-floating-point spectrum?

## X. Conclusion

The current trajectory of AI infrastructure investment proceeds on an unexamined architectural assumption: that the workload pattern of the first generation of generative AI deployments — centralized cloud APIs serving per-token billed inference — will continue to characterize the field indefinitely. This paper has argued that the assumption is mistaken. Inference is fundamentally a different workload from training; it is parallelizable, latency-sensitive, and entirely viable on commodity consumer and workstation hardware. The economic, energy, and policy consequences of separating these workloads are substantial and quantifiable.

We have proposed concrete device class specifications (PAC and CAW) and an eight-component reference architecture that, taken together, constitute the technical specification of an edge-native AI computing ecosystem. Several of the architectural components are subjects of pending patent applications by the first author; they are presented here as open design principles with the intent of contributing to standards work rather than restricting implementation. The patent filings document the inventive priority while the publication establishes the principles in the public technical literature.

Realizing the transition described here is not primarily a hardware problem — the hardware exists. The transition is a software, distribution, and standards problem, and these are tractable on a 5-year horizon. We invite the academic systems community, hardware vendors, AI laboratories, enterprise software organizations, and standards bodies to engage with the architectural framework presented here and to build, criticize, and improve it.

***"Train once. Distribute as software. Infer everywhere. This is the only AI architecture that scales to a billion users without burning a planet's worth of power."***

---

### Acknowledgments

*The authors thank the open-source community contributors to vLLM, llama.cpp, GGUF, GPTQ, and the broader on-device inference ecosystem; their work makes the architectural vision described here practically realizable today. The Apple Machine Learning Research group's open documentation of Apple Intelligence and Private Cloud Compute architecture has been a*

*significant reference point. The first author thanks AMD, Cisco, Intel, and IBM colleagues for many years of formative discussions on the design of high-performance compute infrastructure. Any errors of fact, judgment, or omission are entirely the authors'.*

---

*— END OF DOCUMENT —*

*Preprint · Independent research · 2025 · Vinjam, Narukulla*